\documentclass[
10pt,aps,
notitlepage, 
bibnotes,
prb,
twocolumn,
showpacs,preprintnumbers,superscriptaddress,
amsmath,amssymb,floatfix]{revtex4-1}

\usepackage{graphicx}
\usepackage{dsfont}
\usepackage{dcolumn}
\usepackage{bm}
\usepackage{color}
\usepackage{xcolor}
\usepackage{url}
\usepackage{tabularx}
\usepackage{array}
\usepackage{booktabs}
\usepackage{comment}
\usepackage{hyperref}
\usepackage{footnote}
\usepackage{lipsum}
\usepackage[normalem]{ulem}
\usepackage{booktabs}

\definecolor{colorA}{rgb}{0, 0, 1}
\definecolor{colorB}{rgb}{0.5, 0, 0.9}
\definecolor{colorC}{rgb}{0.4, 0, 0.4}
\definecolor{color_green}{rgb}{0, 0.39, 0}

\makeatletter
  \def\my@tag@font{\normalsize}
  \def\maketag@@@#1{\hbox{\m@th\normalfont\my@tag@font#1}}
  \let\amsmath@eqref\eqref
  \renewcommand\eqref[1]{{\let\my@tag@font\relax\amsmath@eqref{#1}}}
\makeatother

\begin{document}

\title{Skyrmions and antiskyrmions in  monoaxial chiral magnets}

\author{Vladyslav~M.~Kuchkin}
\email{vkuchkin@hi.is}
\affiliation{Science Institute, University of Iceland, 107 Reykjavík, Iceland}
\author{Nikolai~S.~Kiselev}
% \email{n.kiselev@fz-juelich.de}
\affiliation{Peter Gr\"unberg Institute and Institute for Advanced Simulation, Forschungszentrum J\"ulich and JARA, 52425 J\"ulich, Germany}

\date{\today}

\begin{abstract}
We show that competition between local interactions in monoaxial chiral magnets provides the stability of two-dimensional (2D) solitons with identical energies but opposite topological charges.
These skyrmions and antiskyrmions represent metastable states in a wide range of parameters above the transition into the saturated ferromagnetic phase. 
The symmetry of the underlying micromagnetic functional gives rise to soliton zero modes allowing efficient control of their translational movement by the frequency of the circulating external magnetic field.
We also discuss the role of demagnetizing fields in the energy balance between skyrmion and antiskyrmion and in their stability.
\end{abstract}
\maketitle 

% \section{Introduction}

\begin{figure*}[ht]
\centering
\includegraphics[width=17cm]{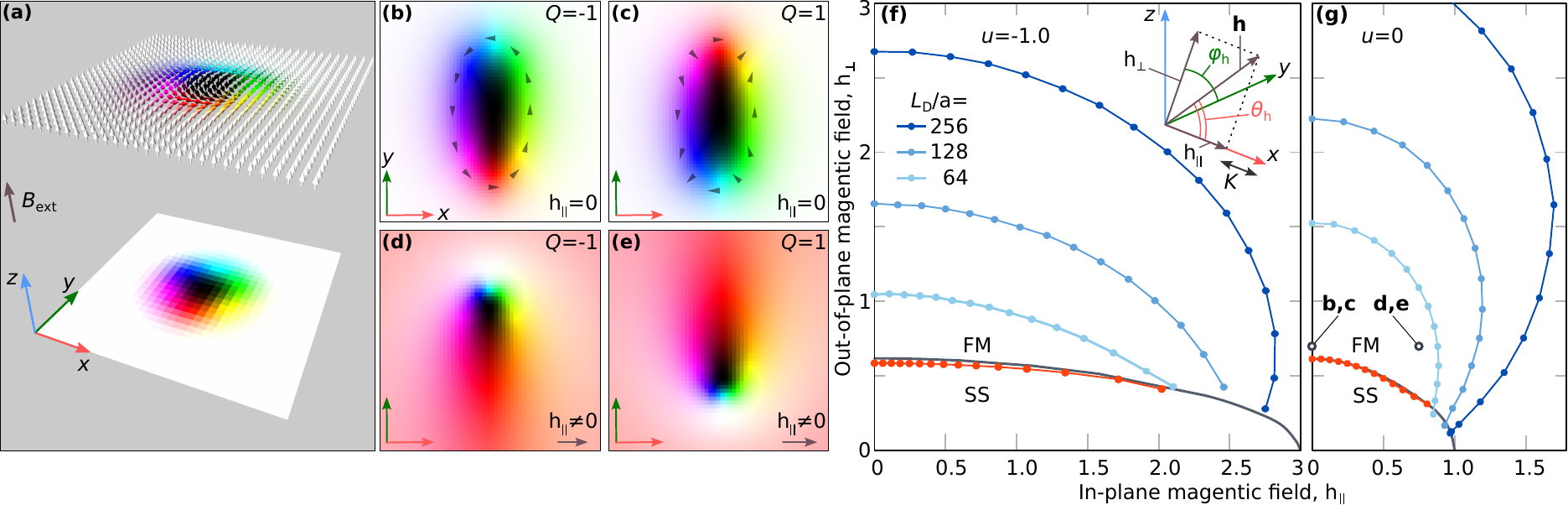}
\caption{~\small \textbf{a} shows axially symmetric skyrmion in isotropic chiral magnet visualized by the vector field (top) and color-coded pixel map (bottom).
\textbf{b}-\textbf{g} correspond to monoaxial chiral magnet case.
\textbf{b} and \textbf{c} show akyrmion antiskyrmion, respectively, in perpendicular field, $h_\perp=0.65$, $h_\parallel=0$, and zero anisotropy, $u=0$.
\textbf{d} and \textbf{e} show skyrmion and antiskyrmion in the tilted magnetic field $h_\perp = h_\parallel = 0.65$ and $u=0$.
The size of window in \textbf{b}-\textbf{e} is $1L_{\mathrm{D}}\times 1L_{\mathrm{D}}$ while entire simulated domain is $4L_{\mathrm{D}}\times4L_{\mathrm{D}}$.
The mesh density in the finite-difference scheme is 64 nodes per $L_{\mathrm{D}}$.
\textbf{f} and \textbf{g} show the phase diagrams $(h_\perp, h_\parallel)$ calculated for easy-plane anisotropy, $u=-1$ and zero anisotropy, $u=0$, respectively. 
The solid black lines correspond to the phase transition between the spin spiral (SS) and the saturated ferromagnetic (FM) states. 
The red curves correspond to the elliptic instability of skyrmion and antiskyrmion.
The blue lines correspond to the collapse field of skyrmion and antiskyrmion, which depends on the mesh discretization density in numerical simulation.
}\label{Fig1}
\end{figure*}

\textit{Introduction.}—Magnetic skyrmions are 2D topological solitons stabilized by the competition of different energy terms of corresponding micromagnetic functional.
In particular, \textit{chiral} magnetic skyrmions are 2D solitons in magnetic crystals with broken inversion symmetry, which in the presence of strong spin-orbit coupling, gives rise to chiral Dzyaloshinskii-Moriya interaction~\cite{Dzyaloshinskii, Moriya} (DMI). 
In the most general case, the ground state of chiral magnets is the spin spiral (SS) state characterized by fixed chirality and the wave vector $\mathbf{q}$ pointing in particular crystallographic directions.
In the particular case of crystals with weak magnetocrystalline anisotropy and isotropic exchange and DMI, \textit{e.g.}, B20-type crystals such as MnSi, FeGe, and FeCoSi, the SSs with $\mathbf{q}$ in different directions are degenerate states.
In the presence of an external magnetic field, isotropic chiral magnets often exhibit a phase transition into the skyrmion phase~\cite{Bogdanov_89, Bogdanov_1994} representing axially symmetric solitons (Fig.~\ref{Fig1}\textbf{a}) arranged in a hexagonal lattice.
Such transitions were experimentally observed in thin films of many B20-type chiral magnets.
Due to the topological nature of such a transition, it typically requires elevated temperatures to overcome the energy barrier associated with the nucleation of skyrmions.
Due to the effect of \textit{chiral surface twist}, which provides an additional energy gain for skyrmion, the thickness of the sample also plays a crucial role.
Above critical thickness, the conical and stacked spiral phases become energetically more favorable compared to skyrmions lattice. 
In this case, skyrmions may appear only in a metastable state, and their experimental observation requires certain efforts.
The same is true for materials with anisotropic DMI~\cite{Hoffmann_17}, where chiral skyrmions lose axial symmetry and usually appear as a metastable state in the whole range of external magnetic fields. 
The most extreme case represents so-called \textit{monoaxial chiral magnets} -- the crystals where DMI is completely vanished or is negligibly small in all except one crystallographic direction. 
Prominent examples of such materials are 
CrNb$_{3}$S$_{6}$,\cite{Togawa_12, Nabei_20, Inui_20,Du_22}, MnNb$_{3}$S$_{6}$~\cite{DiTusa_19,DiTusa_21}, and CrTa$_{3}$S$_{6}$~\cite{Zhang_21}.
The ground state of that system is the SS which some authors called \textit{soliton lattice}.
In an external magnetic field, the SS undergoes  a first-order or second-order phase transition into the saturated ferromagnetic (FM) state depending on the angle between the external magnetic field to the principal axis of the crystal\cite{Ovchinnikov_16}.
All experimental and theoretical studies of these materials show only these two phases.
The existence of skyrmions has never been predicted in this class of magnets.

In this letter, we show that in monoaxial chiral magnets, the skyrmions and their topological counterparts, antiskyrmions, may appear as a metastable state above the critical field corresponding to the transition into the saturated ferromagnetic phase.
In the main part of the work, we discuss a 2D model of a monoaxial chiral magnet where the energies of skyrmion and antiskyrmion turn out to be identical.
Later we show that in the finite thickness plates, considering demagnetizing field effects, this balance is slightly broken in favor of skyrmions.
We also report the unique dynamic properties of skyrmions in monoaxial chiral magnets, particularly their constant velocity motion under the external magnetic field circulating in the plane orthogonal to the principal axis of the crystal.

We have to note that during the preparation of this work, we became aware of the experimental study by Li with co-workers~\cite{Du_22}.
In this pioneering work, the authors report on direct observation of field-induced instability of spin spirals in nanostripes of CrNb$_{3}$S$_{6}$, which leads to the appearance of magnetic skyrmions which in general agrees with our theoretical predictions. 
However, the authors of Ref.~\cite{Du_22} argue that demagnetizing fields play a significant role in the stability of skyrmions in monoaxial materials. 
Here we dispute this statement and show that the mechanism for the stability of skyrmions and antiskyrmions in monoaxial crystals is similar to that in isotropic chiral magnets and governed by the competition between Heisenberg exchange, DMI, and Zeeman energy terms.

% \section{Model}
%
\textit{Model.}—The micromagnetic model of a monoaxial chiral magnet consists of Heisenberg exchange, DMI, and potential energy terms:
\begin{eqnarray}
\mathcal{E}\!=\!\int\! \left[\mathcal{A}\left(\nabla\mathbf{n}\right)^{2} \!+\! \mathcal{D}\!\left(n_\mathrm{z}\dfrac{\partial n_\mathrm{y}}{\partial x}-n_\mathrm{y}\dfrac{\partial n_\mathrm{z}}{\partial x}\right)\!+\!U(\mathbf{n})\right] \mathrm{d}V,
\label{Energy}
\end{eqnarray}
where $\mathbf{n}=\mathbf{M}/M_\mathrm{s}$ magnetization unit vector field, $\mathcal{A}$ is the exchange stiffness constant, and $\mathcal{D}$ is the DMI constant.
We assume that the principal axis of the crystal is parallel to the $x$-axis.
The potential energy term $U(\mathbf{n})$ includes the easy plane anisotropy ($\mathcal{K}<0$), and the interaction with the external magnetic field, $\mathbf{B}_\mathrm{ext}$, and the demagnetizing fields, $\mathbf{B}_\mathrm{s}$, produced by the sample: 
\begin{eqnarray}
    & U(\mathbf{n})= -\mathcal{K}n_\mathrm{x}^{2}-M_\mathrm{s}\left(\mathbf{B}_\mathrm{e}+\mathbf{B}_\mathrm{d}\right)\cdot\mathbf{n} .
    \label{potential}
\end{eqnarray}

Following the standard approach, we introduce 
dimensionless magnetic field $\mathbf{h}=\mathbf{B}_\mathrm{e}/B_\mathrm{D}$,
anisotropy 
$u=\mathcal{K}/(M_\mathrm{s}B_\mathrm{D})$,
and length $r/L_\mathrm{D}$, where
 $L_\mathrm{D}=4\pi\mathcal{A}/\mathcal{D}$ is the equilibrium  period of SS, and $B_\mathrm{D} = \mathcal{D}^{2}/2\mathcal{A}M_\mathrm{s}$ is the saturation magnetic field.
The ground state of the system is SS with $\mathbf{q}\parallel\mathbf{e}_\mathrm{x}$ ($|\mathbf{q}|=2\pi/L_\mathrm{D}$).
It is convenient to parameterize the magnetic field in projections parallel and orthogonal to the  $\mathbf{e}_\mathrm{x}$ as $\mathbf{h}=h_{\parallel}\mathbf{e}_\mathrm{x}+h_{\perp}\mathbf{e}_\mathrm{z}$ or via spherical angles $(\theta_\mathrm{h}, \varphi_\mathrm{h})$ as $\mathbf{h}=h(\cos\theta_\mathrm{h},\sin\theta_\mathrm{h}\cos\varphi_\mathrm{h},\sin\theta_\mathrm{h}\sin\varphi_\mathrm{h})$ (see inset in Fig.~\ref{Fig1}\textbf{f}).

\textit{Skyrmion and antiskyrmion in 2D.}—In the case of a magnetic field perpendicular to the principal axis $h_\parallel=0$, the numerical minimization of the functional \eqref{Energy} in the 2D case gives stable solutions for skyrmion and antiskyrmion (Fig. \ref{Fig1} \textbf{b}, \textbf{c}) characterized by an opposite sign of topological charge
\begin{equation}
Q = \dfrac{1}{4\pi}\int\!  \mathbf{n}\cdot \left(\partial_\mathrm{x}\mathbf{n}\times\partial_\mathrm{y}\mathbf{n}\right) \,\mathrm{d}x\mathrm{d}y.
\label{Qint}
\end{equation}

It is easy to see that the functional \eqref{Energy} is invariant under the $n_\mathrm{x}\mapsto-n_\mathrm{x}$ transformation, $\mathcal{E}(n_\mathrm{x},n_\mathrm{y},n_\mathrm{z})=\mathcal{E}(-n_\mathrm{x},n_\mathrm{y},n_\mathrm{z})$.
Because of that, the energies of skyrmion and antiskyrmion are always identical in this model until we take into account the demagnetizing fields.
One can obtain the equilibrium antiskyrmion from the solution for skyrmion by substituting $n_\mathrm{x}\mapsto-n_\mathrm{x}$, and vice versa, which holds for any tilt of magnetic field (Fig.~\ref{Fig1} \textbf{d} and \textbf{e}).

\begin{figure*}[t]
\centering
\includegraphics[width=17cm]{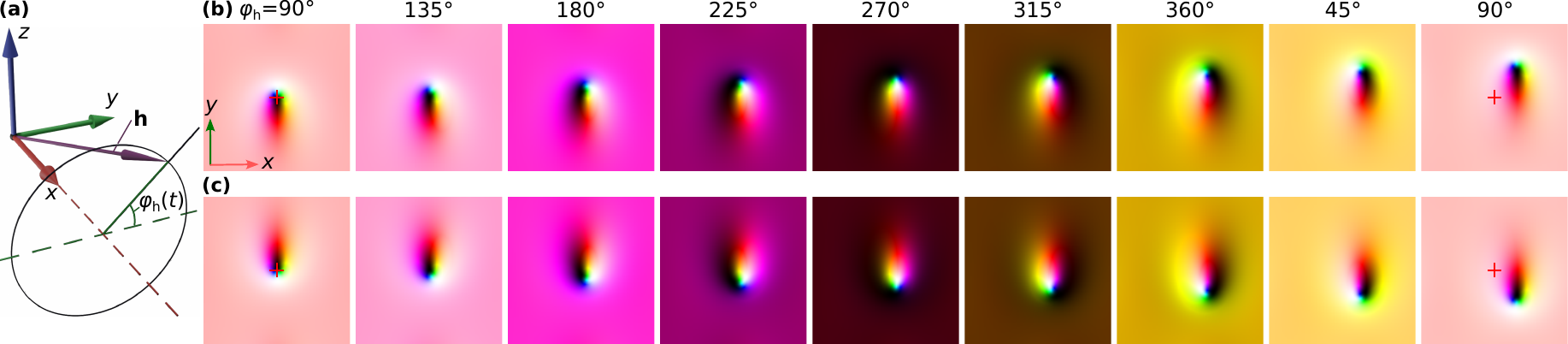}
\caption{~\small Skyrmion dynamics in monoaxial chiral magnet under magnetic field circulating about the principal axis.
\textbf{a} schematic representation of the numerical experiment setup. The principal axis is parallel to $x$, and an external magnetic field $\mathbf{h}$ circulates about it.
The two rows of images in \textbf{b} and \textbf{c} are the snapshots of the skyrmion and antiskyrmion, respectively.
Each snapshot corresponds to the different phase of the field, $\varphi_\mathrm{h}=2\pi\omega t$, circulating with the frequency $\omega=-0.25\mathrm{GHz}$
The value of the reduced field is $h=0.55$, and the tilt angle of the field to the $x$-axis is $\vartheta_\mathrm{h}=\pi/4$.
The red crosses indicate the center of the $4L_{\mathrm{D}}\times4L_{\mathrm{D}}$ simulated domain.
}
\label{Fig3}
\end{figure*}

To illustrate a wide range of skyrmion and antiskyrmion stability in this model, we calculated the phase diagrams, Fig.~\ref{Fig1} \textbf{f} and \textbf{g}, for strong easy-plane anisotropy, $u=-1$, and zero anisotropy, $u=0$, respectively.
In agreement with the previous studies~\cite{Ovchinnikov_16}, the phase diagram consists of two phases only: the SS state at a low magnetic field and the FM state at a high field.
The skyrmion stability range is bounded by the elliptical instability field from below (red curve) and the collapse field from above (blue curve).
The elliptical instability field lies very close to the phase transition line but does not fully coincide with it.
Similar to the model of an isotropic chiral magnet~\cite{Kuchkin_20i}, with increasing mesh density and effectively approaching the micromagnetic limit, the range of skyrmions stability expands due to the increase of the collapse field.
The elliptical instability field, on the other hand, is almost insensitive to mesh discretization density. 
Since both diagrams in Fig.~\ref{Fig1} \textbf{f} and \textbf{g} are qualitatively identical, we deduce that anisotropy is not an essential energy term for skyrmion stability but must be taken into account only for quantitative agreement with experimental data.

It is worth emphasizing that contrary to the isotropic case, the skyrmions and antiskyrmions in a monoaxial material are always metastable solutions with energies higher than the energy of the FM state.
Because of that, the experimental observation of skyrmions in such systems requires special conditions.

A remarkable property of functional \eqref{Energy} is that all terms except demagnetizing fields energy and Zeeman energy are invariant with respect to rotations of the whole magnetic texture about the $x$-axis by arbitrarily angle $\varphi_\mathrm{s}$.
If we ignore, for now, demagnetizing field effects, it is easy to see that for $\varphi_\mathrm{s}=\varphi_\mathrm{h}$, the total energy in \eqref{Energy} remains conserved,
\begin{equation}
\mathcal{E}(\mathcal{R}_x(\varphi_\mathrm{h})\mathbf{n})=\mathrm{const},
\label{Econst}
\end{equation}
here $\mathcal{R}_x(\varphi)$ is the rotation matrix about $x$-axis by angle $\varphi$. 
Thereby the phase diagrams in Fig.~\ref{Fig1} \textbf{f} and \textbf{g} are valid for any $\varphi_\mathrm{h}$.

\begin{figure*}[t]
\centering
\includegraphics[width=17cm]{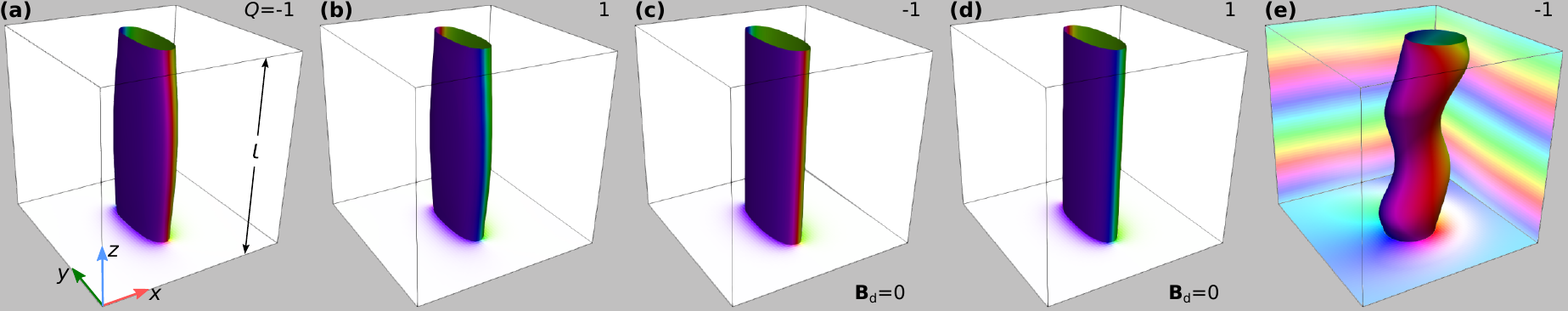}
\caption{~\small Skyrmion and antiskyrmion tubes in a plate of monoaxial chiral magnets calculated with and without demagnetizing fields ($\mathbf{B}_\mathrm{d}=0$).
\textbf{a} and \textbf{b} show skyrmion and antiskyrmion, respectively, calculated with the demagnetizing fields at $B_\mathrm{e}=260$ mT.
\textbf{c} and \textbf{d} show skyrmion and antiskyrmion, respectively, calculated without the demagnetizing fields at $B_\mathrm{e}=97$ mT.
\textbf{e} is the skyrmion tube in an isotropic chiral magnet at $B_\mathrm{e}=200$ mT.
The system size in all calculations was $L_\mathrm{x}=L_\mathrm{y}=4L_\mathrm{D}$, $l=2L_\mathrm{D}$, $L_\mathrm{D}=64a$ with periodic along $xy$ and open in $z$ boundary conditions.
For illustrative purposes, here we show only a fraction of the simulated domain of the size $L_\mathrm{x}=L_\mathrm{y}=2L_\mathrm{D}$. 
}
\label{Fig4}
\end{figure*}

\textit{Skyrmion dynamics.}—Taking into account the above invariance of the skyrmion energy with respect to arbitrary rotation about the principal axis \eqref{Econst}, we consider the skyrmion dynamics under circulating magnetic field~\cite{Ovchinnikov_12}, $\varphi_\mathrm{h}(t)=2\pi\omega t$, $\vartheta_\mathrm{h}=\mathrm{const}$. 
First, we parameterize magnetization $\mathbf{n}$ in a  frame $x^\prime, y^\prime, z^\prime$ related to the frame $x, y, z$ by rotation abut the $x$-axis by angle $\varphi_\mathrm{h}(t)$:
\begin{equation}
\mathbf{n}=\left(\begin{array}{c}
\sin\Theta\cos\Phi\\
\sin\Theta\sin\Phi\cos\varphi_\mathrm{h}(t)-\cos\Theta\sin\varphi_\mathrm{h}(t)\\
\cos\Theta\cos\varphi_\mathrm{h}(t)+\sin\Theta\sin\Phi\sin\varphi_\mathrm{h}(t)\\
\end{array}\right),
\label{TF}
\end{equation}
where $\Theta$ and $\Phi$ are the standard spherical angles in the $x, y, z$ frame. 
Using \eqref{TF}, the LLG equation~\cite{Landau_Lifshitz} can be written as
\begin{eqnarray}
    & \dfrac{\partial\Theta}{\partial t}\sin\Theta=-\dfrac{\gamma}{M_\mathrm{s}}\dfrac{\delta\mathcal{E}}{\delta\Phi}-\alpha\sin^{2}\Theta\dfrac{\partial\Phi}{\partial t}  +f(\omega),\nonumber\\
    & \dfrac{\partial\Phi}{\partial t}\sin\Theta=\dfrac{\gamma}{M_\mathrm{s}}\dfrac{\delta\mathcal{E}}{\delta\Theta}+\alpha\dfrac{\partial\Theta}{\partial t}+g(\omega), \label{LLG_TF}
\end{eqnarray}
where $f(\omega)=2\pi\omega\sin\Theta\left(\alpha\cos\Theta\cos\Phi+\sin\Phi\right)$ and $g(\omega)=2\pi\omega\left(\cos\Phi\cos\Theta-\alpha\sin\Phi\right)$ 
are the torques appearing due to the presence of a time-dependent magnetic field.
The LLG equation in the form \eqref{LLG_TF} is useful to derive the Thiele equation for skyrmion motion.
In particular, assuming a translational motion of the magnetic texture with the constant velocity $\mathbf{v}$, we can write 
\begin{equation}
    \dfrac{\partial\Theta}{\partial t}=-(\mathbf{v}\cdot\nabla)\Theta,\, \dfrac{\partial\Phi}{\partial t}=-(\mathbf{v}\cdot\nabla)\Phi.\label{time_v}
\end{equation}
The criterion of the constant velocity reads~\cite{Malozemoff_79}, \begin{equation}
\displaystyle\int\left(\frac{\delta\mathcal{E}}{\delta\Theta}\nabla\Theta+\frac{\delta\mathcal{E}}{\delta\Phi}\nabla\Phi\right)\mathrm{d}x\mathrm{d}y=0.
\label{v=const}
\end{equation} 
The physical meaning of integral \eqref{v=const} is the dissipation energy~\cite{Malozemoff_79}.
Substituting \eqref{time_v} into \eqref{LLG_TF} and performing the integration in \eqref{v=const}, we obtain the Thiele equation \cite{Thiele_73}:
\begin{equation}
\mathbf{G}\times\mathbf{v}+\alpha\hat{\Gamma}\mathbf{v}=\alpha\mathbf{F},\label{Thiele}
\end{equation}
where $\mathbf{G}= Q\mathbf{e}_\mathrm{z}$ is the gyrovector, $\hat{\Gamma}$ is the dissipation tensor with components $\Gamma_{i j}=(4\pi)^{-1}\int(\partial_{i}\mathbf{n}\cdot\partial_{j}\mathbf{n})\mathrm{d}x\mathrm{d}y$ and $\mathbf{F}=\frac{\omega}{2}\int \left(n_\mathrm{z}\nabla n_\mathrm{y}-n_\mathrm{y}\nabla n_\mathrm{z}\right)\mathrm{d}x\mathrm{d}y$ is the effective force.
The general solution of \eqref{Thiele} has the following form: 
\begin{equation}
    v_\mathrm{x} = \alpha^{2}\dfrac{ F_\mathrm{x}\Gamma_\mathrm{yy} + F_\mathrm{y}Q_{-}}{Q^{2}+\alpha^{2}\Gamma}, \, v_\mathrm{y} = \alpha^{2}\dfrac{ F_\mathrm{y}\Gamma_\mathrm{xx}- F_\mathrm{x}Q_{+}}{Q^{2}+\alpha^{2}\Gamma}, 
    \label{velocity_theory}
\end{equation}
where $\Gamma=\Gamma_\mathrm{xx}\Gamma_\mathrm{yy}-\Gamma_\mathrm{xy}^{2}$, $Q_{\pm}=Q/\alpha\pm\Gamma_\mathrm{xy}$. It is common to refer to the term $\mathbf{F}$ in \eqref{Thiele} as an effective force because it exhibits similarities to forces encountered in classical mechanics within non-inertial reference frames. However, it is essential to note that the Thiele equation describes motion without acceleration, setting it apart from classical mechanics. Therefore, while drawing such analogies, it is necessary to be careful and recognize their limitations.

As follows from \eqref{velocity_theory} the solutions with non-zero velocity exist only for $\alpha>0$ and $|\omega|>0$.
Moreover, for small damping, $\alpha\ll 1$, the velocity components are $\sim \alpha\omega L_\mathrm{D}$, which yield an efficient approach to control the skyrmion velocity and the direction of its motion with the circulating field frequency.

It is worth highlighting the derived Thile equation describes skyrmion dynamics induced by a time-dependent magnetic field which is uniform in space. 
Previous works discussing the role of the time-dependent magnetic field in skyrmion motion~\cite{Fangohr_15, Moon_16, Mochizuki_18} considered the effect of exciting skyrmion eigenmodes and their coupling to translational motion. 
Thus, the maximal velocity was reached when the AC field frequency coincided with the resonance frequencies.
In that case, the derivation of the Thiele equation required introducing averaged over oscillation period magnetization, which can move as a rigid object.
This approach worked only for a small amplitude of the AC field. 
Contrary to that, the derived Thiele equation \eqref{Thiele} can describe skyrmion motion caused by a rotating magnetic field of any amplitude within the skyrmion stability range. 
Skyrmion velocity, governed by this equation, turned out to be proportional to the rotating field frequency. 
Therefore, the described effect has nothing in common with the skyrmion motion driven by oscillating fields.

To verify the Thiele equation \ref{velocity_theory}, we have performed LLG simulations in Mumax code \cite{Mumax} with the following parameters $h_{\parallel}=h_{\perp}=0.55$, $u=0$, $\alpha=1$ and $\omega=250$ MHz.
The snapshots of skyrmion and antiskyrmion at various moments in time are shown in Fig.~\ref{Fig3}. 
Tracing the soliton position with taking into account the periodic boundary conditions in the plane~\cite{Kuchkin_21}, we estimated the velocities and the deflection angle $\beta=\arctan v_\mathrm{y}/v_\mathrm{x} = \pm52.45^\circ$ for skyrmion $(+)$ and antiksyrmion $(-)$, respectively.
Using $\mathbf{n}$-field corresponding to the static solutions of skyrmion and antiskyrmion, according to \eqref{velocity_theory}, the deflection angles are slightly smaller than that in the numerical experiment and equals $\beta=\pm50.1^\circ$.
Such a mismatch occurs due to some deviation of the skyrmion shape in the dynamical regime from the skyrmion shape in statics. 
Using the snapshots of $\mathbf{n}$-field obtained in LLG simulations at steady motion ($t\gg0$), equation \eqref{velocity_theory} gives velocity components nearly identical to that in the numerical experiment, and deflection angle, $\beta=\pm52.46^\circ$.

We also noticed that at $\vartheta_\mathrm{h}=0$, the change in the shape of skyrmions is negligibly small.
Moreover, as follows from \eqref{velocity_theory} for small damping, $\alpha\ll 1$, the velocity components $\sim \alpha\omega L_\mathrm{D}$.
This suggests a simple way to estimate the $\alpha$ parameter experimentally.

It is worth noting that the presence of demagnetizing fields brakes the symmetry of the functional \eqref{Energy} and destroys the zero modes \eqref{Econst}. In this case, the analysis of the skyrmion dynamics becomes more complicated, but the effect of skyrmion motion induced by the circulating magnetic field remains present.
In particular, this influences the stability ranges for solitons, so the dependence on the phase angle $\phi_\mathrm{h}$ can not be excluded.
Skyrmion motion can also be achieved in this case. One must modify both the magnetic field's amplitude and tilt to remain within the skyrmion stability range.

\textit{Skyrmions and antiskyrmions in 3D.}—All presented above effects remain present in the 3D case of finite thickness plates and taking into account the demagnetizing field effects. To illustrate this, we performed micromagnetic simulations in an extended plate of a monoaxial chiral magnet with material parameters for CrNb$_{3}$S$_{6}$, $\mathcal{A}=0.733$ pJ, $L_\mathrm{D}=48$ nm, $M_\mathrm{s}=155$ kA/m and $\mathcal{K}=-146$ kJ/m$^{3}$, taken from Ref.\cite{Du_22}.
The mumax-script is provided in Supplemental materials.
Skyrmion and antiskyrmion tubes stabilized at $B_\mathrm{e}=260$ mT are shown in Fig.~\ref{Fig4} \textbf{a} and \textbf{b}. Small modulations close to the open surface are present.
The calculated Lorentz TEM images for skyrmion and antiskyrmion provided in the Supplemental materials demonstrated a hardly seen difference. Thus, to experimentally distinguish skyrmion and antiskyrmion tubes in these materials requires additional analysis, for instance, based on the dynamical properties of solitons.

To investigate the role of the demagnetizing field in solitons' stabilization, we performed micromagnetic simulations in the same system, ignoring demagnetizing field presence.
In this case, we can obtain stable skyrmion and antiskyrmion in the field $B_\mathrm{z}=97$ mT (Fig.~\ref{Fig4} \textbf{c}, \textbf{d}).
The shape of these solitons is similar to those shown in Fig.~\ref{Fig4}.
Thus, the demagnetizing fields are not a key ingredient in skyrmion's stabilization in monoaxial chiral magnets.
For comparison with the isotropic case, in Fig.~\ref{Fig4} \textbf{e}, we show a stable skyrmion tube embedded in the conical phase background, where we use the same material parameters but assuming an isotropic DMI and $\mathcal{K}=0$.

\textit{Conclusions.}—In summary, we have investigated the properties of monoaxial chiral magnets and discovered stable solutions for both skyrmions and antiskyrmions in 2D and 3D systems.
We calculated stability ranges for skyrmions in 2D systems at the tilted magnetic field with varying $L_\mathrm{D}$ values, which are relevant for both spin and micromagnetic models.
Additionally, we have demonstrated the feasibility of moving skyrmions and antiskyrmions in these systems using a rotating external magnetic field, employing both analytic Thiele approaches and micromagnetic simulations. 
In the case of thin films of monoaxial chiral magnets, we have confirmed the coexistence of skyrmions and antiskyrmions.
Based on the micromagnetic simulations for material CrNb$_{3}$S$_{6}$,  we found physical parameters at which skyrmion and antiskyrmion tubes can be experimentally observed.

VMK is grateful to A. N. Bogdanov for valuable discussions.
The authors acknowledge financial support from the Icelandic Research Fund (Grant No. 217750).
This project has received funding from the European Research Council under the European Union's Horizon 2020 Research and Innovation Programme (Grant No.~856538 - project ``3D MAGiC'').

% \appendix
% \onecolumngrid

\end{document}